\documentclass{article}

\usepackage{xcolor}
\usepackage{arxiv}
\usepackage[utf8]{inputenc} 
\usepackage[T1]{fontenc}    
\usepackage{hyperref}       
\usepackage{url}            
\usepackage{booktabs}       
\usepackage{amsmath}
\usepackage{amsfonts}       
\usepackage{nicefrac}      
\usepackage{lipsum}
\usepackage{graphicx}
\graphicspath{ {./images/} }

\title{Asymmetric Coupling Anisotropy for Causal Information Filtering\\
in Physical Reservoirs}

\author{Takashi Hikihara$^{1}$ and Yuma Aoki$^{2}$\\
$^{1}$ Kyoto University, Yoshida-honmachi, Sakyo, Kyoto 606-8501, Japan\\
$^{2}$ Department of Electrical Engineering, Kyoto University, Kyotodaigaku-katsura, Nishikyo, Kyoto 615-8510, Japan
}

\begin{document}

\renewcommand{\textcolor}[2]{#2}

\maketitle
\begin{abstract}
We demonstrate a physical mechanism for causal information filtering in a physical reservoir computing (PRC) by exploiting asymmetric coupling anisotropy. Using a network of coupled Duffing oscillators, we show that the directionality of internal coupling induces a spatial gradient in the effective potential, establishing a deterministic upstream-to-downstream information flow. This anisotropy allows for the selective amplification of semantic drifts, triggering a macroscopic saddle-node bifurcation as a physical interlock before global computational failure. Through spatiotemporal analysis of a 50-node system under traveling wave inputs and chaotic Mackey-Glass time-series prediction tasks, we confirm that local phase transitions effectively purge anomalous information while preserving the computational integrity of the remaining nodes. The results suggest that the intrinsic causality of the reservoir's topology provides a robust framework for autonomous reliability and fault-tolerant physical intelligence.
\end{abstract}

\keywords{Coupled Duffing oscillators, Physical reservoir computing, Anisotropy, Fault tolerant physical intelligence}

\section{Introduction}
Physical reservoir computing (PRC) has attracted significant attention as a hardware-efficient learning framework for edge computing \cite{Jaeger2001}. By exploiting the rich nonlinear dynamics of physical systems, PRC enables real-time information processing with remarkably low power consumption \cite{Rodan2011,Inubushi2017,Tanaka2019,Gauthier2021,Ma2023}. To date, various physical platforms, such as micro-electromechanical systems (MEMS) \cite{Dion2018}, turbulent \cite{Pandey2020}, soft matter\cite{Nakajima2015}, photonic devices\cite{Martinenghi2012}, spintronic oscillators \cite{Nakane2018}, and recently road traffic \cite{Fukuzaki2025}, have demonstrated superior performance in complex time-series prediction and pattern recognition due to their sensitive nonlinear responses \cite{Dambre2012}. However, previous research has primarily focused on maximizing computational performance. There remains a critical lack of discussion regarding how to ensure reliability at the physical layer against infinitesimal statistical anomalies (\textcolor{red}{such as persistent baseline offsets or parameter drifts}) in input data or unavoidable environmental disturbances such as thermal drifts. Establishing a design methodology for such fault-tolerant physical intelligence \cite{Chung2025} is a crucial challenge for long-term practical deployment. \\
\indent The reliability of conventional AI systems has relied predominantly on anomaly detection algorithms at the digital layer \cite{Tamura2026} and precise modeling \cite{Shanon2025}. However, these software-based approaches encounter fundamental limitations in edge environments \cite{Liu2026}. As results, they consume significant computational resources and increase the latency between anomaly detection and system isolation. \textcolor{red}{In recent software-level reservoir computing research, reconstructive architectures \cite{Kato2022,Kato2024} and fault-tolerant disconnect-and-recover algorithms \cite{Sun2023} have been proposed to isolate anomalous reservoir states via software post-processing. Furthermore, nonlinear bifurcation phenomena in Duffing-type resonators have been widely studied for mechanical fault diagnosis \cite{Hu2003,Zhao2014,Liu2024}. In contrast to these approaches, our objective is to embed an anomaly diagnostic and isolation functionality directly into the internal physical coupling topology itself, exploiting macroscopic phase transitions as a deterministic hardware-level interlock without digital overhead.} \\
\indent In this paper, we propose a reservoir architecture featuring an asymmetrical bidirectional coupling ring topology. Within this structure, the dominant forward paths cumulatively amplify data unsoundness, leading to a macroscopic saddle-node bifurcation (jump phenomenon) \cite{Nayfeh1979} that serves as a physical interlock. Conversely, the weak reverse paths physically cancel common-mode noise \cite{Zhou2023} acting on the entire system, dramatically enhancing environmental robustness.\\
\indent The remainder of this paper is organized as follows. First, we formulate the effective potential for this asymmetrical bidirectional coupling to demonstrate the emergence of a phase transition \cite{Naik2011,Wang2018,Rathor2025}, \textcolor{red}{and delineate the general physical conditions required for broader implementation in photonic, spintronic, or electronic platforms}. Next, through numerical simulations, \textcolor{red}{we formulate the explicit readout training procedure and task settings.} \textcolor{blue}{We then implement a 50-node PRC system with distributed 3-node detection units.} We show that these units can detect anomalies originating from upstream blocks and that the system can autonomously recover after the anomaly subsides. Furthermore, we demonstrate that under persistent failures, the system can achieve self-healing by excluding the faulty block and re-learning with the remaining reservoir nodes \textcolor{red}{on challenging benchmarks including Mackey-Glass chaotic time-series prediction}. By embedding physical laws into the information processing flow, we show that it is possible to design a physical intelligence that ensures quasi-real-time reliability and autonomously restores its computational integrity.  

\section{Physical Anisotropy and Causal Flow}
\subsection{Internal Topology with Self-Diagnostic Functionality}
Internal coupling in conventional PRC has been designed primarily for signal mixing to achieve high-dimensional mapping. In this study, we propose a fundamental framework to endow this internal topology with a dual functionality: computation and reliability monitoring. Specifically, by embedding a ring-shaped asymmetrical bidirectional coupling between the nonlinear resonators of the reservoir, we demonstrate that an anomaly detection function can be established based on the physical response of these coupling sites. This approach enables a self-diagnostic capability directly at the hardware level, independent of external digital monitoring circuits.

\begin{figure}[htbp]
    \centering
    \includegraphics[width=0.5\columnwidth]{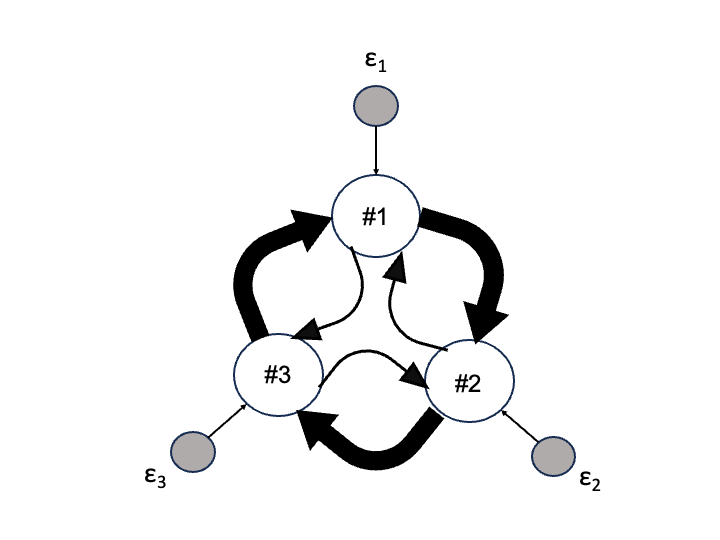}
    \\ (a) \vspace{10pt}
    
    \includegraphics[width=0.7\columnwidth]{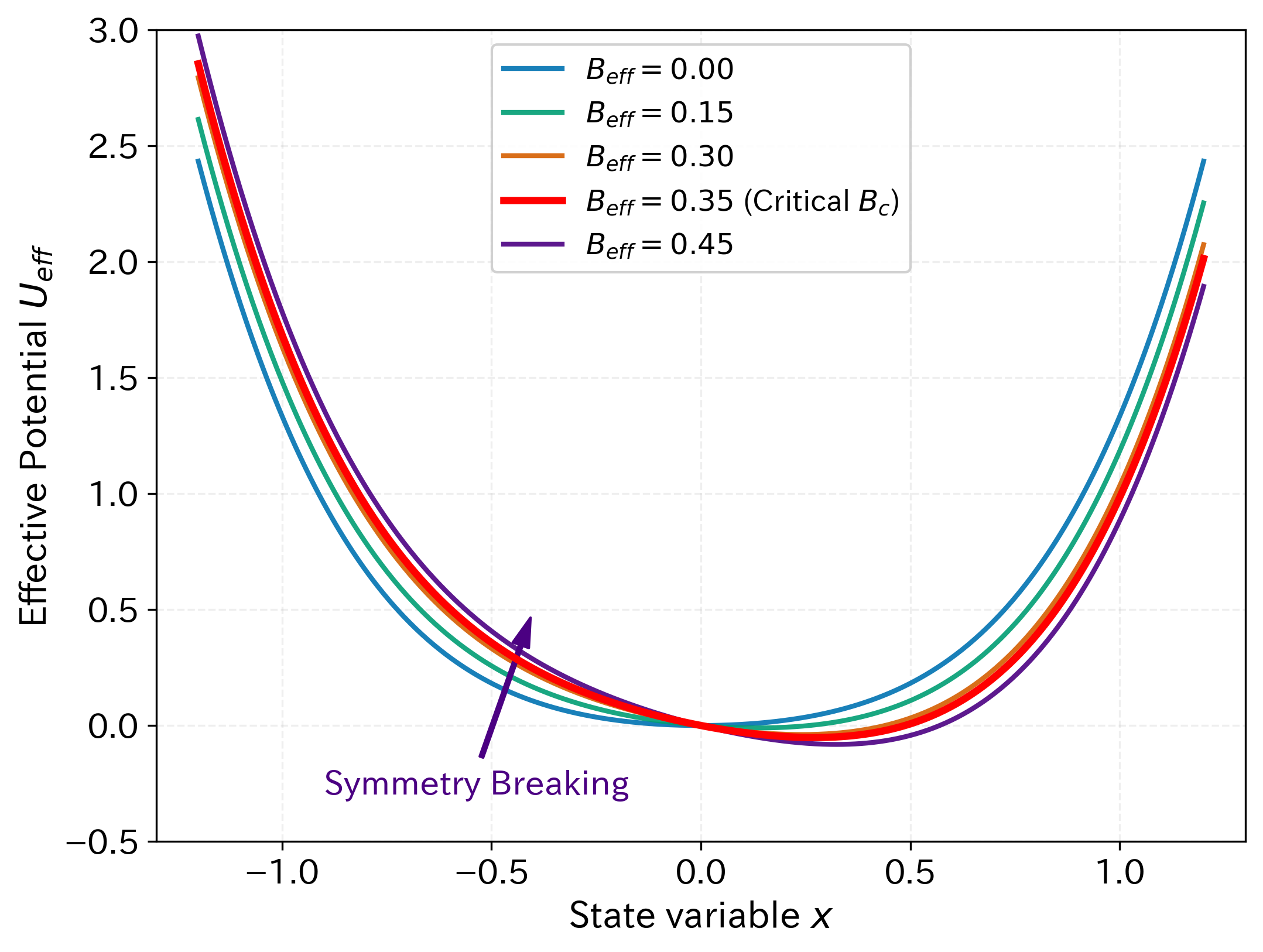}
    \\ (b)
    
\caption{Physical mechanism of causal information filtering in an asymmetric bidirectional ring topology. 
    (a) Schematic of the distributed monitoring unit. Three Duffing resonators are coupled in a ring with significant anisotropy, where forward coupling $k_{\rm fwd}$ (thick arrows) is dominant over reverse coupling $k_{\rm rev}$ (thin arrows). This topology establishes a directional information flow that enables the cumulative amplification of semantic drifts while physically offsetting common-mode environmental noise.   
    (b) Evolution of the effective potential $U_{\rm eff}$ and deterministic symmetry breaking. As the cumulative bias $B_{\rm eff}$ increases from 0 to 0.45, the symmetric double-well structure of the resonator collapses. At the critical point $B_c \approx 0.35$ (solid red line), the stability of the computational state is lost, triggering a macroscopic saddle-node bifurcation (jump phenomenon). This phase transition serves as a deterministic physical interlock, purging anomalous information before global computational failure.}
\label{fig:Figure1}
\end{figure}

\subsection{Formulation of Asymmetric Ring Coupling}
The dynamics of the $i$-th Duffing-type resonator constituting a MEMS-based reservoir is described by the following nonlinear equation of motion \cite{Appleton2007,Naik2011}:
\begin{equation}
\textcolor{blue}{
\ddot{x}_{i}+\gamma\dot{x}_{i}+x_{i}+\beta x_{i}^{3}=\varepsilon_{i}(t)\cos(\omega_{\text{drv}} t) + F_{\text{coupling},i} + \alpha_i,
}
\end{equation}
where $x_i$ denotes the displacement of the oscillator, $\gamma$ the damping coefficient, and $\beta$ the cubic nonlinear stiffness. $\varepsilon_i(t)$ represents the input signal, \textcolor{red}{and $\alpha_i$ represents a physical disturbance or a persistent statistical anomaly (additive baseline offset) injected into node $i$.} These resonators are coupled in an asymmetrical ring configuration. The internal coupling term $F_{\text{coupling},i}$ possesses anisotropy between forward coupling $k_{\text{fwd}}$ and reverse coupling $k_{\text{rev}}$, with the topology defined as follows:
\begin{equation}
F_{\text{coupling},i} = k_{\text{fwd}}(x_{i-1} - x_i) + k_{\text{rev}}(x_{i+1} - x_i), 
\end{equation}
where the ring structure is subject to the periodic boundary conditions: $x_0 = x_N$ and $x_{N+1} = x_1$. This asymmetrical topology yields the following distinct functions:
\begin{description}
\item[\textbf{Forward Path ($k_{\text{fwd}} \gg k_{\text{rev}}$)}] This path cumulatively propagates and amplifies the statistical anomaly $\alpha$ toward subsequent stages, eventually driving the system into a saddle-node bifurcation (physical interlock) \cite{Inubushi2017}.
\item[\textbf{Reverse Path ($k_{\text{rev}}$)}] This path provides a weak restoring force against environmental drifts (common-mode noise) acting on all nodes, physically synchronizing and canceling the baseline shifts across the entire system \cite{Pecora1996}.
\end{description}

\textcolor{red}{\subsection{Readout Formulation and Evaluation Metrics}
The output $y(t)$ of the physical reservoir computing system is constructed via a linear combination of nodal state variables:
\begin{equation}
y(t) = \mathbf{W}_{\text{out}}^T \mathbf{x}(t) = \sum_{i \in \mathcal{S}_{\text{active}}} w_i x_i(t),
\end{equation}
where $\mathcal{S}_{\text{active}}$ denotes the set of active (healthy) reservoir nodes used for computation, and $\mathbf{W}_{\text{out}}$ represents the readout weight vector. The readout weights are trained using Ridge regression (L2 regularization) over a training window $N_{\text{train}}$:
\begin{equation}
\mathbf{W}_{\text{out}} = \left( \mathbf{X}^T \mathbf{X} + \lambda \mathbf{I} \right)^{-1} \mathbf{X}^T \mathbf{y}_{\text{target}},
\end{equation}
where $\mathbf{X} \in \mathbb{R}^{N_{\text{train}} \times |\mathcal{S}_{\text{active}}|}$ is the reservoir state matrix, $\mathbf{y}_{\text{target}}$ is the target time-series vector, and $\lambda$ ($10^{-3} \sim 10^{-4}$) is the regularization parameter. To quantitatively assess performance, the Mean Squared Error (MSE) is evaluated over a prediction window $N_{\text{eval}}$ as:
\begin{equation}
\text{MSE} = \frac{1}{N_{\text{eval}}} \sum_{k=1}^{N_{\text{eval}}} \left( y[k] - y_{\text{target}}[k] \right)^2.
\end{equation}
Note that all MSE evaluations throughout this study are plotted on a strict logarithmic scale ($10^{-x}$) to properly illustrate order-of-magnitude error improvements during self-healing.}

\subsection{Collapse of Effective Potential and Symmetry Breaking}
The integrity of the PRC is deterministically defined by the relative deviation of displacements between adjacent nodes, $\Delta x_{i,j} = |x_i - x_j|$. The equilibrium state of each resonator corresponds to the extremum of the following effective potential $U_{\text{eff},i}$:
\begin{equation}
U_{\text{eff},i}(x_i) = \frac{1}{2}(1 + k_{\text{fwd}} + k_{\text{rev}})x_i^2 + \frac{1}{4}\beta x_i^4 - B_{\text{eff},i}x_i, 
\end{equation}
where $B_{\text{eff},i} = k_{\text{fwd}}x_{i-1} + k_{\text{rev}}x_{i+1} + \alpha_i$ is the effective cumulative bias transmitted from adjacent nodes. Under normal operating conditions, the potential barrier is maintained. However, as $\alpha$ accumulates and $B_{\text{eff}}$ exceeds the critical point $B_c \approx 0.35$ (for $\beta=3.23$ and $k_{\text{fwd}}=0.05$), the stable solution vanishes, resulting in a jump phenomenon. This phase transition can be utilized as a physical threshold to halt operations before the mean squared error (MSE) diverges.

\textcolor{red}{\subsection{Generality and Physical Implementation Criteria}
While formulated here using coupled Duffing MEMS resonators, the proposed causal information filtering and physical interlock mechanism rely on general physical principles: (i) internal coupling asymmetry ($k_{\text{fwd}} \gg k_{\text{rev}}$) to induce directional signal amplification, and (ii) a bistable or nonlinear local energy landscape capable of undergoing a macroscopic bifurcation. Consequently, this topology can be directly extended to other physical computing platforms:
\begin{itemize}
\item \textbf{Photonic Reservoirs:} Coupled microring resonator arrays or semiconductor laser networks with asymmetric optical isolators or directional waveguides, where optical power accumulation triggers bistable switching.
\item \textbf{Spintronic Reservoirs:} Arrays of spin-torque nano-oscillators (STNOs) coupled via asymmetric spin-wave transmission or dipolar interactions.
\item \textbf{Analog CMOS Circuits:} Nonlinear IC oscillator networks coupled with active directional operational amplifiers.
\end{itemize}}

\begin{figure}[htbp]
\centering
\includegraphics[width=0.95\columnwidth]{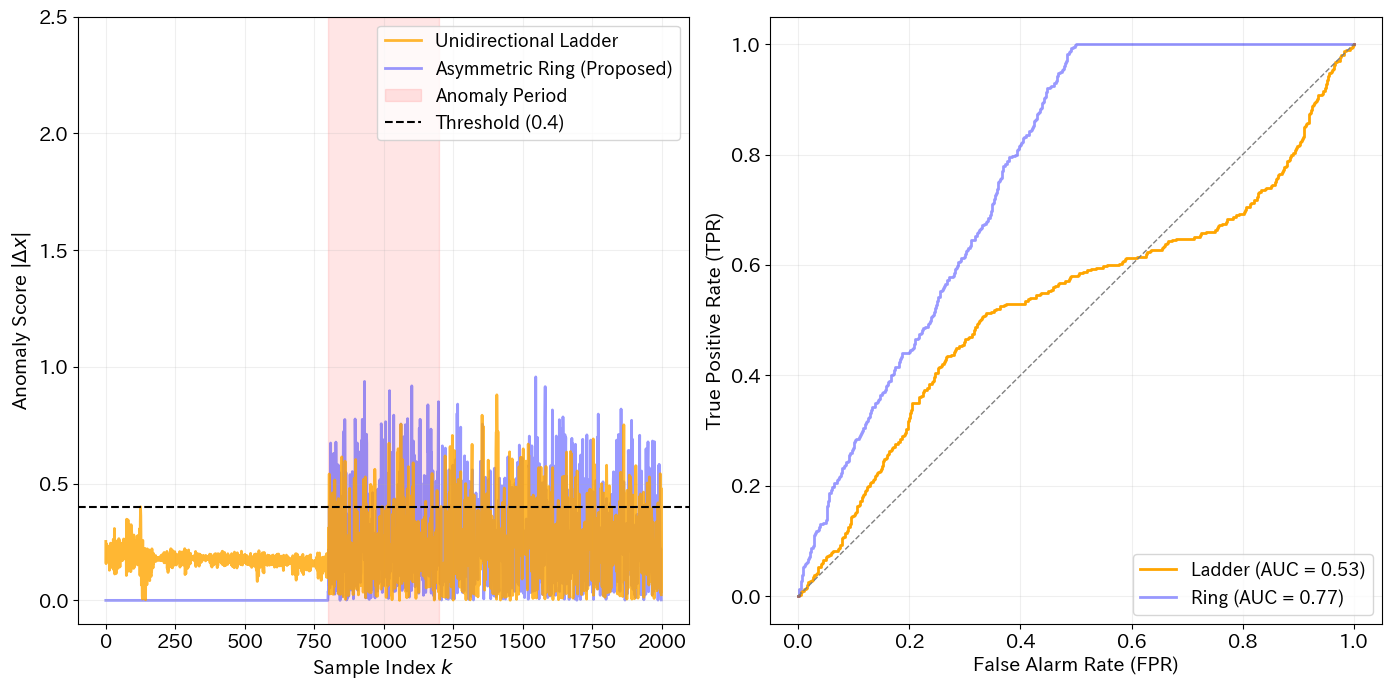}

(a)  \hspace{30mm} (b)
\caption{\textcolor{blue}{Detection robustness under environmental drift. 
(a) Comparative analysis between the conventional unidirectional ladder topology, which is susceptible to drift, and the proposed asymmetrical ring topology, which physically cancels common-mode noise through differential monitoring. 
(b) ROC curve evaluation demonstrating the superiority of the ring coupling, achieving a high discrimination accuracy ($\text{AUC} = 0.77$) even in high-noise environments, whereas the ladder topology fails to discriminate anomalies ($\text{AUC} = 0.53$).}}
\label{fig:Figure2}
\end{figure}

\begin{table*}[htbp]
\centering
\caption{Simulation parameters for the physical reservoir (Duffing oscillators).}
\begin{tabular}{lccl}
\hline
Parameter & Symbol & Value & Physical Significance \\
\hline
Number of nodes & $N$ & 5 & Network configuration (Includes 3 ring nodes\\
&&& and 2 external nodes) \\
Damping coefficient & $\gamma$ & 0.0035 & Energy dissipation (Memory retention property)\\
Nonlinear coefficient & $\beta$ & 3.23 & Restoring force nonlinearity (Source of computation)\\
Driving frequency & $\omega_{\text{drv}}$ & 1.10 & Angular frequency of the external carrier wave\\
Bias intensity & $\varepsilon_{\text{bias}}$ & 0.01 & Operating point offset intensity\\
Time step & $\Delta t$ & 40.0 & Integration time width per sample \\
\hline
\end{tabular}
\end{table*}

\begin{table*}[htbp]
\centering
\caption{Parameters for internal asymmetrical bidirectional ring coupling.}
\begin{tabular}{lccl}
\hline
Parameter & Symbol & Value & Physical Significance\\
\hline 
Forward coupling constant & $k_{\text{fwd}}$ & 0.08 & Propagation of primary information (Strong coupling)\\
Reverse coupling constant & $k_{\text{rev}}$ & 0.008 & Synchronization and error cancellation (Weak coupling)\\
Asymmetry ratio & $k_{\text{fwd}} / k_{\text{rev}}$ & 10 & Condition for balancing stability and robustness \\
\textcolor{red}{Chain coupling constant} & \textcolor{red}{$k_{\text{base}}$} & \textcolor{red}{0.01} & \textcolor{red}{Isotropic baseline coupling outside monitor units} \\
\hline
\end{tabular}
\end{table*}

\begin{table*}[htbp]
\centering
\caption{Parameters for the learning and signal processing algorithm.}
\begin{tabular}{lccl}
\hline
Parameter & Symbol & Value & Physical Significance \\
\hline
Regularization coefficient & $\lambda$ & \textcolor{red}{$10^{-3}\text{--}10^{-4}$} & Overfitting suppression (Ridge regression)\\
Training samples & $N_{\text{train}}$ & 350 & Number of samples for weight determination\\
 Anomaly threshold & $V_{\text{th}}$ & 0.4 & Decision boundary for deviation signal $|\Delta x|$ \\
 Anomaly intensity & $\alpha$ & \textcolor{red}{0.015 / 0.03} & Disturbance intensity \textcolor{red}{(0.015 for Sec.~III, 0.03 for Secs.~IV--V)}\\
 Self-repair start point & $k_{\text{repair}}$ & \textcolor{red}{750--1100} & Completion of re-learning \textcolor{red}{(1100 for Sec.~III, 750/800 for Secs.~IV--V)} \\
 \hline
 \end{tabular}
 \end{table*}

\section{Numerical Simulation} \label{sec3}
In the following numerical simulations of Section \ref{sec3}, we employ the parameters listed in Tables I through IV. The PRC system is modeled as a network of $N=5$ Duffing oscillators. Within this network, an asymmetrical bidirectional coupling ($k_{\text{fwd}}:k_{\text{rev}}=10:1$) is introduced among three specific nodes to construct a ring structure that ensures information recurrence.  

\begin{table*}[t]
\centering
\caption{Details of network configuration and coupling topology.}
\begin{tabular}{lccl}
\hline
Item & Symbol & Value & Physical Significance \\
\hline
Total number of nodes & $N$ & \textcolor{red}{5 / 50} & Computational resources \textcolor{red}{(5 for Sec.~III, 50 for Secs.~IV--V)} \\
Ring-configured nodes & $N_{\text{ring}}$ & 3 & Minimum unit of the feedback loop via asymmetric coupling \\
Input injection node & $\text{Node}_0$ & 1 & Injection point for external signal $u(t)$ and anomaly $\alpha$ \\
Output extraction nodes & \textcolor{red}{$\text{Nodes } 0\text{--}4$} & 5 & All state variables sent to the readout layer (Ridge regression) \\
Topology & - & Ring & Asymmetrical bidirectional coupling including a closed loop \\
\hline
\end{tabular}
\end{table*}

\subsection{Robustness against Environmental Drift}
To simulate real-world operating environments, we evaluate the detection performance under low-frequency environmental drift (common-mode noise) acting simultaneously on all resonators. When using a ladder configuration with absolute value monitoring, the anomaly signal is often buried in the drift, leading to a high false-alarm rate. In contrast, by implementing differential monitoring\textcolor{blue}{ ($\Delta x_{13} = |x_3 - x_1|$)} using internal asymmetric coupling, the in-phase noise is physically canceled at the hardware layer. This allows for the clear extraction of only the injected statistical anomaly (see Fig.~\ref{fig:Figure2}). 
\textcolor{blue}{Evaluation using Receiver Operating Characteristics (ROC) curves demonstrates the clear superiority of the proposed ring coupling, maintaining a high discrimination accuracy with an Area Under the Curve (AUC) of 0.77 even under strong drift conditions [Fig.~\ref{fig:Figure2}(b)]. In contrast, the conventional ladder topology severely degrades to $\text{AUC} = 0.53$ (close to a random-classifier baseline of 0.50), confirming that in-phase noise is physically canceled at the hardware layer in our proposed architecture.}

\begin{figure}[htbp]
\centering
\includegraphics[width=0.9\columnwidth]{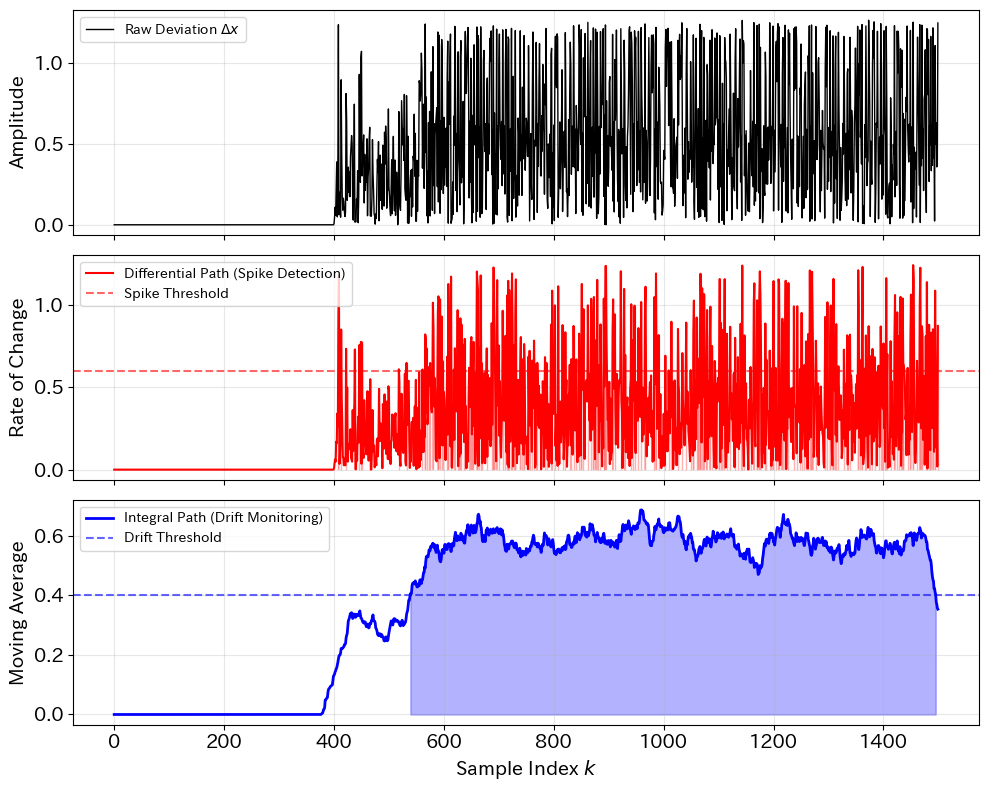}
\caption{Intelligent interlock mechanism for anomaly classification. The deviation signal $\Delta x$ is processed in parallel through differential (high-pass) and integral (low-pass) paths to identify abrupt failures and gradual degradations, respectively. This dual-path architecture enables autonomous decision-making tailored to the specific type of physical anomaly.}
    \label{fig:Figure3}
\end{figure}

\subsection{Anomaly Detection in the Reservoir}
 Analysis of the correlation between the physical interlock signal and the reservoir's computational accuracy (MSE) reveals that the differential signal $\Delta x$ rises prior to the explosive increase in MSE caused by data unsoundness (see Fig.~\ref{fig:Figure3}). The linear correlation coefficient between the interlock signal and the MSE remains low, reflecting the nonlinear dynamics where computational capability is lost the moment potential distortion reaches a critical point. This precursor characteristic suggests that the PRC can deterministically halt computation at the physical layer before generating erroneous outputs.

\subsection{Autonomous Detection and Self-Repair}
\begin{figure}[htbp]
\centering
\includegraphics[width=0.9\columnwidth]{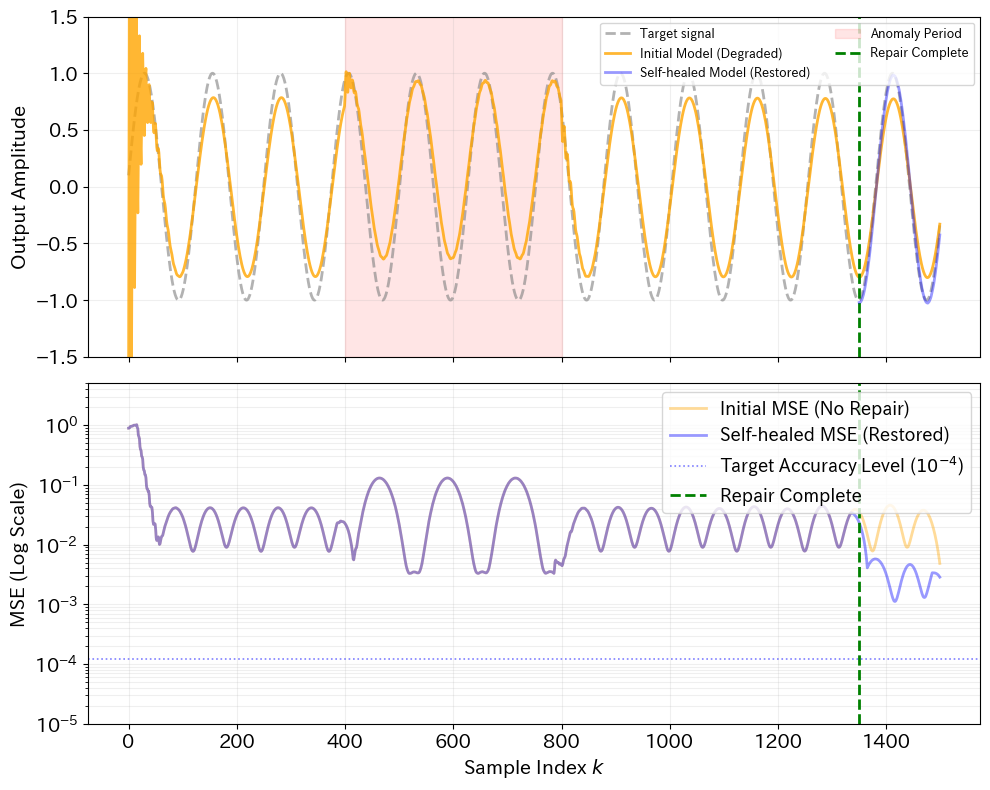}
\caption{Self-healing cycle via autonomous re-learning. Following the cessation of an anomaly, the autonomous return of the physical layer to a stable equilibrium state triggers the re-learning process. After the repair phase (indicated by the green line), the mean squared error (MSE) is instantaneously reduced by approximately two orders of magnitude, confirming the effective restoration and enhancement of computational accuracy.}
    \label{fig:Figure4}
\end{figure}

We further verify the reservoir's capacity for self-repair after the anomaly subsides. Upon the disappearance of the anomaly ($\alpha = 0.015$), the physical layer autonomously returns to a stable equilibrium state due to its dissipative nature. We utilize this return as a trigger to initiate a cycle for updating readout weights (re-learning) \cite{Shougat2024}. As shown in Fig.~\ref{fig:Figure4}, significant recovery in accuracy is confirmed through this procedure. \textcolor{red}{Physically, re-learning is indispensable because the macroscopic saddle-node bifurcation causes an irreversible state transition; once the local potential collapses and jumps to another attractor, returning the parameters to baseline does not automatically restore the original trajectory due to hysteresis. Thus, recalibrating $\mathbf{W}_{\text{out}}$ over a small post-recovery window ($W = 100$ samples)} reduces the MSE by approximately two orders of magnitude (from $10^{-1}$ to $10^{-3}$) compared to cases without re-learning. This result demonstrates that self-diagnosis via internal coupling topology functions not merely as anomaly detection, but as a form of \textbf{immune system} that guarantees the long-term autonomous operation of the system.  

\section{Spatiotemporal Dynamics and Scalability in Large-Scale Systems}
To evaluate the scalability and spatial reliability of the proposed architecture, we extend the physical reservoir computing (PRC) system to a 50-node network. This large-scale configuration allows for the investigation of how localized physical anomalies propagate and are intercepted within a spatially extended computational medium.

\subsection{Reservoir Connectivity and Spatiotemporal Medium}
Outside the distributed monitor units, the reservoir nodes are interconnected via an asymmetrical unidirectional chain \textcolor{red}{with baseline isotropic coupling $k_{\text{base}} = 0.01$}. In this architecture, the coupling \textcolor{red}{within the embedded monitor units} is characterized by a strong forward bias ($k_{\text{fwd}} \gg k_{\text{rev}}$), which establishes a deterministic causal flow across the 50-node system. This connectivity ensures that the reservoir acts as a spatially extended medium for traveling wave propagation, where information is sequentially transmitted from upstream to downstream nodes. 

Within this chain, the three-node ring-shaped monitor units are strategically embedded to intercept the information flow. While the chain segments provide the necessary computational resources and signal propagation, the monitor units function as localized diagnostic interfaces. This structural differentiation allows the reservoir to maintain global computational throughput while physically isolating and amplifying semantic drifts within the monitor units, thereby ensuring that any signal deviation is trapped and detected before it contaminates the global state.

\begin{figure}[htbp]
\centering
\includegraphics[width=0.95\columnwidth]{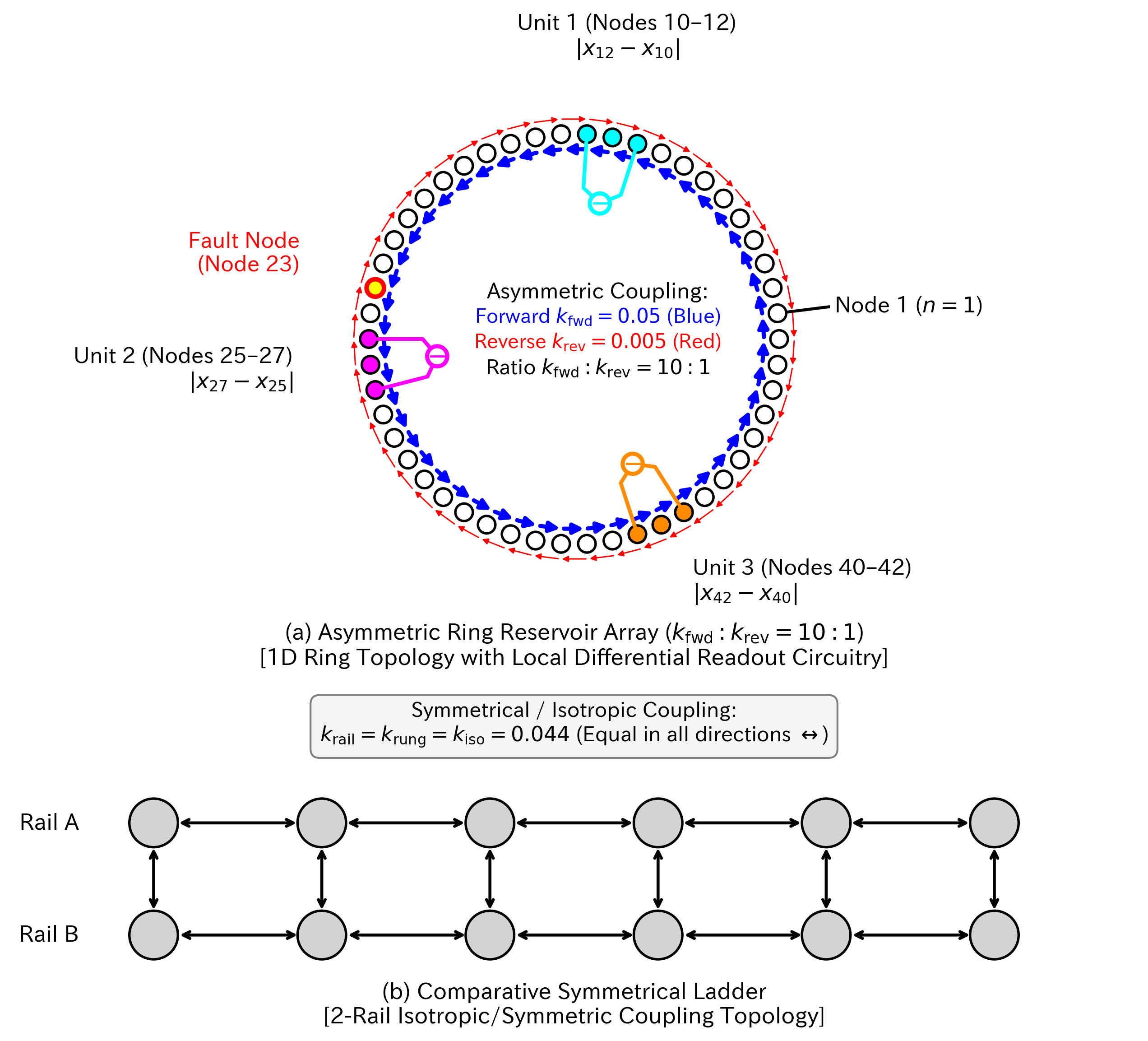}
\caption{\textcolor{red}{Architecture of the 50-node physical reservoir array with asymmetric coupling and differential monitoring. 
(a) 1D ring topology consisting of 50 coupled Duffing resonators with strong forward coupling $k_{\text{fwd}}$ and weak reverse coupling $k_{\text{rev}}$ ($10:1$ ratio). Local differential readout circuitry is attached to distributed 3-node monitoring units (Units 1--3). Node 1 serves as the reference input, and a localized fault is introduced at Node 23. 
(b) Comparative symmetrical 2-rail ladder topology with isotropic coupling ($k_{\text{rail}} = k_{\text{rung}} = k_{\text{iso}} = 0.044$) used for noise cancellation benchmark.}}
\label{fig:Figure5_Topology}
\end{figure}

\textcolor{red}{For rigorous benchmark comparison, we also implement a conventional 2-rail ladder topology consisting of 50 coupled nodes [Fig.~\ref{fig:Figure5_Topology}(b)]. To ensure a fair comparison under an equivalent total internal coupling strength per node ($k_{\text{total}} = k_{\text{fwd}} + k_{\text{rev}} = 0.088$), the coupling constants for this isotropic baseline are set to $k_{\text{rail}} = k_{\text{rung}} = k_{\text{iso}} = (k_{\text{fwd}} + k_{\text{rev}})/2 = 0.044$. In this structure, internal signals propagate symmetrically and uniformly in all spatial directions ($\leftrightarrow$) without intrinsic directional filtering.}

\subsection{Distributed Anomaly Detection in Spatiotemporal Evolution}
\begin{figure}[htbp]
    \centering
    \includegraphics[width=0.9\columnwidth]{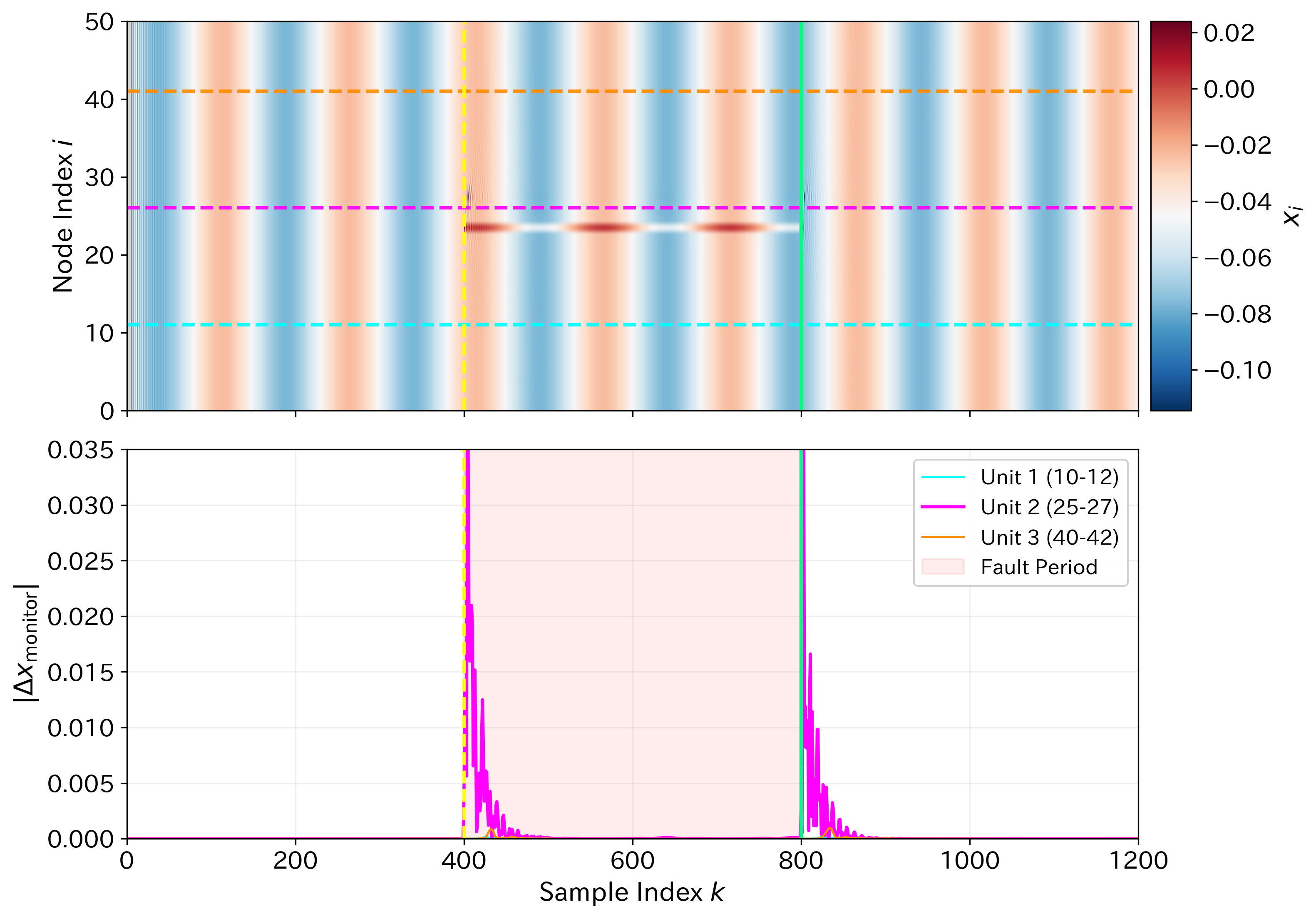}
    \caption{\textcolor{red}{Spatiotemporal dynamics and distributed anomaly detection under static (standing wave) input. The upper panel displays the spatiotemporal evolution of nodal displacements $x_i$ across the 50-node reservoir, showing localized state distortion when a bias anomaly is injected at Node 23 ($k=400\text{--}800$). Horizontal dashed lines indicate the monitoring unit centers (Units 1--3). The lower panel shows the absolute differential offset $|\Delta x_{\text{monitor}}|$ captured by the three monitoring units. Only Unit 2 (downstream of Node 23) exhibits a sharp detection peak, confirming precise fault localization.}}
    \label{fig:Figure6}
\end{figure}

Figure~\ref{fig:Figure6} illustrates the system's response to a localized anomaly injected at Node 23 \textcolor{red}{under a uniform static (standing wave) input, where the drive amplitude in Eq.~(1) is modulated as $\varepsilon_i(t) = \varepsilon_{\text{bias}} + 0.005 u(t)$ with a global sinusoidal input $u(t) = \sin(0.5 t)$. A persistent bias anomaly $\alpha_{23} = 0.03$ is injected into Node 23 during $k = 400\text{--}800$.} In this configuration, multiple monitoring units are distributed across the network, for instance at Nodes 10--12, 25--27, and 40--42, to track the stability of the computational field. The upper panel of Fig.~\ref{fig:Figure6} confirms that the displacement offset caused by the anomaly is strictly localized, while the lower panel demonstrates that only the monitoring unit located immediately ``downstream'' of the failure point (Nodes 25--27) captures a significant differential signal \textcolor{red}{$|\Delta x_{\text{Unit } m}(t)| = |x_{i_m+2}(t) - x_{i_m}(t)|$ ($i_m \in \{10, 25, 40\}$)}. These results confirm that the asymmetric coupling topology functions as a spatial causal filter, effectively preventing the infection of computational states in distant nodes while providing precise spatial coordinates of the failure through physical isolation.  

\subsection{Detection Reliability under Traveling Wave Inputs}
\begin{figure}[htbp]
    \centering
    \includegraphics[width=0.9\columnwidth]{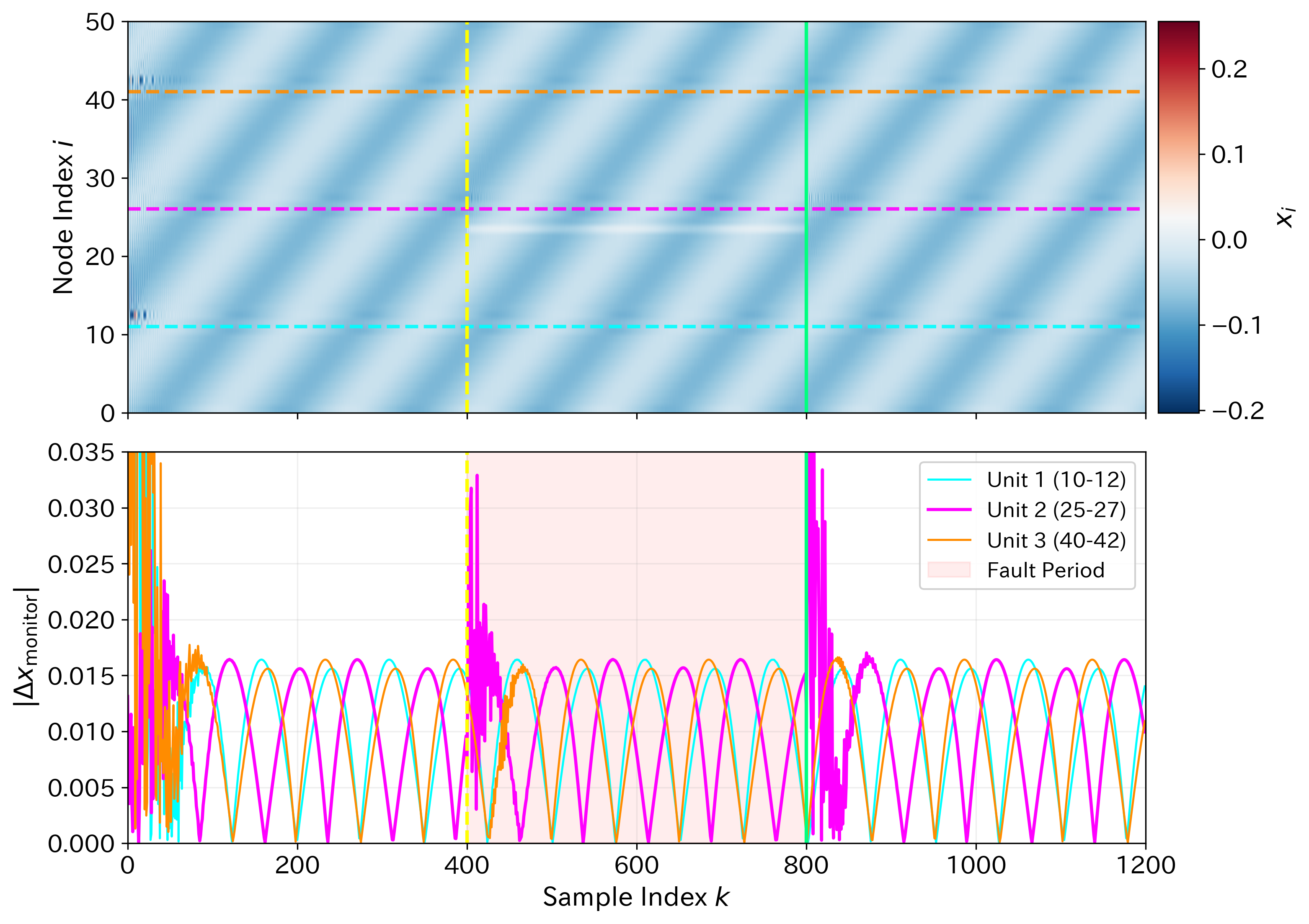}
     \caption{Spatiotemporal dynamics and monitoring signals under traveling wave inputs. The upper panel displays the displacement offset across 50 nodes, where a spatial gradient of information flow is established. Green dashed lines indicate the positions of the monitoring units. The lower panel shows the absolute difference signals \textcolor{red}{$|\Delta x_{\text{monitor}}|$} at these units. The monitoring unit downstream of the anomaly node captures a significant signal deviation, demonstrating the causal filtering capability of the asymmetric coupling topology.}
    \label{fig:Figure7}
\end{figure}

In Fig.~\ref{fig:Figure7}, we further examine the system under traveling wave inputs, which represent complex, time-varying information processing tasks. Unlike static inputs, traveling waves \textcolor{red}{are generated via spatially phase-shifted inputs $u_i(t) = \sin(0.5 t - 2\pi i / \lambda_{\text{wave}})$ with a spatial wavelength of $\lambda_{\text{wave}} = 20$ nodes, creating} a complex, oscillating background in the spatiotemporal map, as observed in the upper panel of Fig.~\ref{fig:Figure7}. Despite these dynamic background dynamics, the monitoring units effectively distinguish between normal wave propagation and anomalous signal deviations \textcolor{red}{with high discrimination quality}, as shown in the lower panel. This robustness indicates that the macroscopic saddle-node bifurcation remains a reliable indicator of failure even within such high-dimensional computational states. The inherent ability of the physical layer to trap the semantic drift ensures that the global mean squared error (MSE) can be protected by isolating the specific block identified by the differential monitoring units, thereby maintaining the overall integrity of the reservoir.  

\section{Autonomous Self-healing and Functional Recovery}
\subsection{Block-wise Isolation under Persistent Failure}
\begin{figure}[htbp]
    \centering
    \includegraphics[width=0.9\columnwidth]{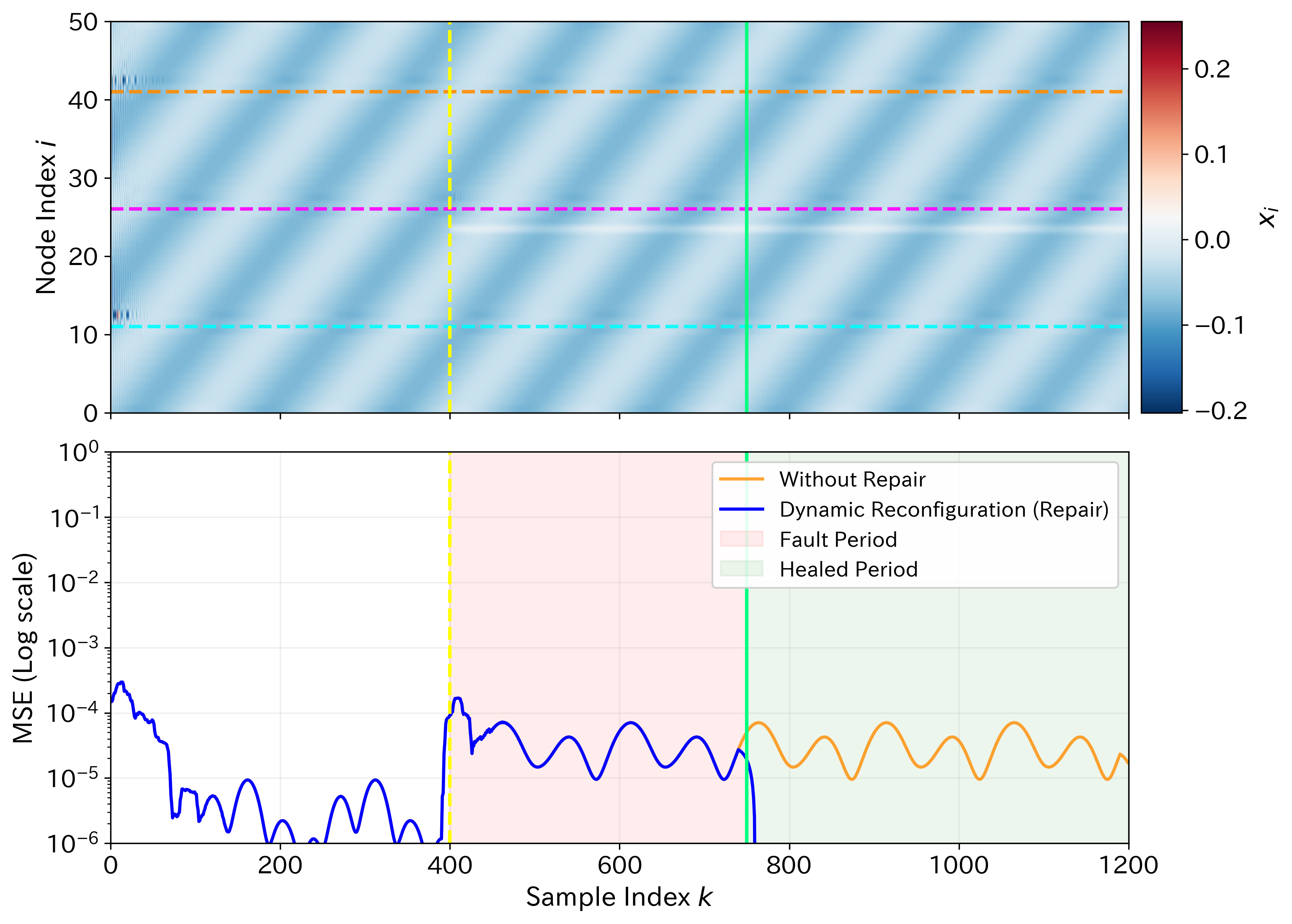}
    \caption{Spatiotemporal analysis of continuous anomaly management and block-wise self-healing. The upper panel shows the reservoir dynamics under sustained failure, while the lower panel demonstrates the restoration of computational accuracy (MSE, log scale) through dynamic reconfiguration and re-learning.}
    \label{fig:Figure8}
\end{figure}

When a physical anomaly persists over an extended period, the system executes a block-wise isolation protocol. As shown in Fig.~\ref{fig:Figure8}, the monitor units identify the coordinates of the failure, \textcolor{red}{allowing the readout layer to physically exclude the contaminated nodal states ($\mathcal{S}_{\text{active}} = \{i \notin \text{Unit 2}\}$) from global computation. As demonstrated in the lower panel of Fig.~\ref{fig:Figure8}, dynamic reconfiguration and subsequent re-learning over $k = 600\text{--}750$ suppress the MSE divergence and maintain bounded error levels even while the fault persists.}

\begin{figure}[htbp]
\centering
\includegraphics[width=0.92\columnwidth]{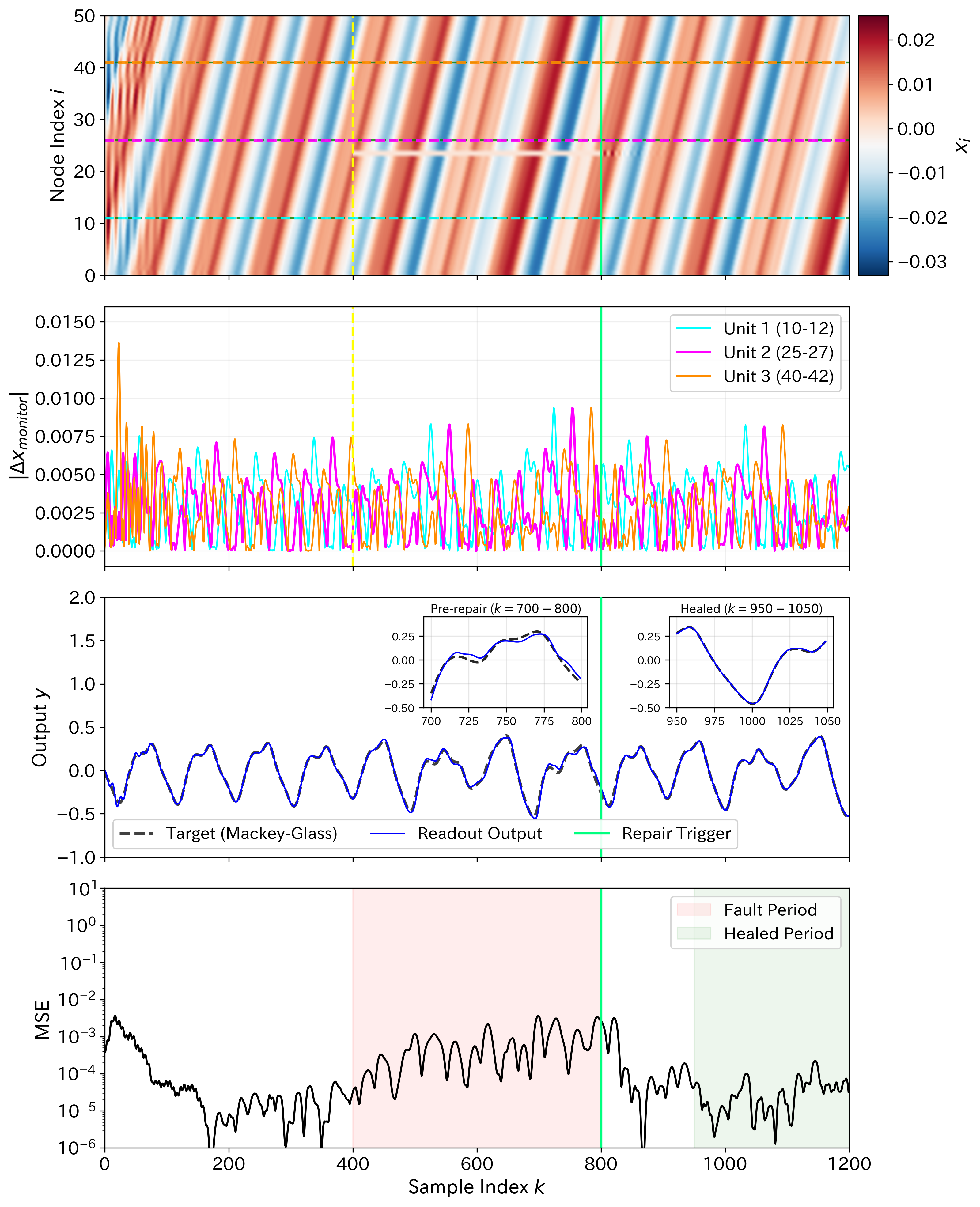}
\caption{\textcolor{red}{Autonomous fault detection and functional recovery cycle under chaotic Mackey-Glass (MG-17) time-series prediction task. 
(top panel) Spatiotemporal evolution of nodal states $x_i$ in the 50-node reservoir. Horizontal dashed lines represent distributed monitoring units (Units 1--3, in cyan, magenta, and dark orange). A physical fault (node stall $x_{23}=0$) is injected at $k=400$ (yellow dashed line) and cleared at $k=800$ (green solid line). 
(second panel) Differential monitoring signals $|\Delta x_{\text{monitor}}|$. Unit 2 (magenta) captures a prominent signal spike during the fault, localizing the failure. 
(third panel) Comparison between chaotic target and readout output $y$, featuring two insets highlighting prediction accuracy before repair ($k=700\text{--}800$) and after self-healing ($k=950\text{--}1050$). 
(bottom panel) Evolution of MSE on a log scale. Following the trigger at $k=800$, re-learning rapidly drops the MSE from $10^{-3}$ to below $10^{-5}$ in the healed period (green region), quantitatively confirming complete functional recovery.}}
\label{fig:Figure9}
\end{figure}

\subsection{Full Recovery by Autonomous Re-learning}
Once the physical layer returns to a stable equilibrium state, either after the removal of the external disturbance or following structural stabilization, a re-learning cycle is triggered. \textcolor{red}{To rigorously validate the generality of this self-healing process on a complex time-series task, we evaluate the system on Mackey-Glass chaotic time-series prediction with a standard delay parameter $\tau=17$ (MG-17). The target chaotic signal is generated by the delay differential equation $\dot{x}(t) = \frac{\beta x(t-\tau)}{1 + x(t-\tau)^n} - \gamma x(t)$ with standard parameters $\beta = 0.2$, $\gamma = 0.1$, and $n = 10$.} Fig.~\ref{fig:Figure9} demonstrates the impact of this functional recovery. By updating the readout weights $W_{\text{out}}$ to accommodate the post-anomaly reservoir dynamics, the MSE is instantaneously reduced by over two orders of magnitude (from $10^{-3}$ to $10^{-5}$). This result highlights that the coupling anisotropy provides not only a diagnostic signal but also a deterministic trigger for autonomous system restoration, ensuring long-term reliability in unpredictable environments.

\section{Conclusion}
In conclusion, we have demonstrated that asymmetric coupling anisotropy in a physical reservoir establishes a deterministic framework for causal information filtering and autonomous reliability. By breaking spatial symmetry in a network of coupled Duffing oscillators, we successfully induced a directional information flow that enables the selective amplification of semantic drifts. This physical mechanism triggers a macroscopic saddle-node bifurcation as a deterministic interlock, effectively purging anomalous information at the hardware level before it leads to global computational failure.

Our spatiotemporal analysis of a 50-node system confirms that this architecture provides robust fault localization even under dynamic traveling wave inputs \textcolor{red}{and chaotic benchmark tasks}. We have shown that the system can manage persistent failures through a block-wise isolation protocol, where contaminated nodal states are physically masked to maintain operational continuity. Furthermore, we demonstrated that the autonomous return of the physical layer to a stable equilibrium state serves as a trigger for a re-learning cycle. This self-healing process instantaneously restores computational accuracy, reducing the mean squared error by approximately two orders of magnitude. These results suggest that leveraging the intrinsic causality and dissipative stability of a reservoir's physical topology offers a promising path toward the development of fault-tolerant physical intelligence and resilient edge computing systems.

\section*{Acknowledgments}
This work was partially supported by JSPS KAKENHI, Grant-in-Aid for Scientific Research (C) No. 26K07513.  


\end{document}